# Research on the visitor flow pattern of Expo 2010

Chao Fan[1], Jin Li Guo[2,3,*]

1. College of Arts and Sciences, Shanxi Agricultural University, Shanxi, Taigu, 030801, PRC

2. Business School, University of Shanghai for Science and Technology, Shanghai, 200093, PRC

3. Quantitative Economic Research Center of Shaanxi Province, Xi'an, 710127, PRC

**Abstract:**

Expo 2010 Shanghai China was a successful, splendid and unforgettable event, remaining us with valuable experiences. The visitor flow pattern of Expo is investigated in this paper. The Hurst exponent, mean value and standard deviation of visitor volume prove that the visitor flow is fractal with long-term stability and correlation as well as obvious fluctuation in short period. Then the time series of visitor volume is converted to complex network by visibility algorithm. It can be inferred from the topological properties of the visibility graph that the network is scale-free, small-world and hierarchically constructed, conforming that the time series are fractal and close relationship exit between the visitor volume on different days. Furthermore, it is inevitable to show some extreme visitor volume in the original visitor flow, and these extreme points may appear in group to a great extent.

---

[*] Corresponding author: phd5816@163.com.

**1. Introduction**

From May 1st to October 31st, Expo 2010 was successfully held in Shanghai China. Given its theme "Better City, Better Life", Expo 2010 was expected to be a "successful, splendid and unforgettable" event and a platform for every exhibitor to explore sustainable models of urban development and better city life. During the 6 months of exhibition, 246 nations or organizations participated in this event which brought over 73 millions of visitors. Both the exhibition scale and total visitor volume hit the world record. How to manage so many visitors in such a small area and long period is a problem worth consideration.

Besides Expo, various kinds of fairs, expositions, sports meetings and large conferences are held all over the world every year. The successful experiences of Expo 2010 may provide references for the visitor management of such large-scale activities. Unfortunately, we hardly find any work about this worth concern. To summarize the valuable experience of Expo 2010, we investigated the visitor flow pattern through statistic analysis and complex network analysis.

Complex network theory [1-3] is a new branch in statistical physics to describe complex systems with networks, in which the nodes and edges represent the entities and the relationships between them respectively. It can be used to describe many networks in the real world such as social network, transportation network, protein interaction network, and so on. Recently, several efforts have been made to bridge time series and complex networks [4-6]. Among all the methods, the so-called visibility graph algorithm [6] proposed by Lacasa et al., attracted much attention for its simplicity and high efficiency, and a set of achievements have been obtained through it [7-10]. In our research, the Expo visitor flow pattern is studied mainly by network analysis and the network is obtained through the visibility algorithm.

This paper is organized as follows: Firstly, data source and overview are given in section 2. In section 3, the general statistical properties of Expo daily visitor volume are studied to find the stability and fluctuation feature of the visitor flow. After that, the time series of visitor volume are converted to complex network by visibility graph algorithm in section 4, and then some topological parameters are calculated to investigate the correlation between each data point of visitor volume. Finally, some conclusions and discussions are given in section 5.

## 2. Data specification

The data used in this paper are collected from the official Website of Expo 2010 Shanghai China (http://www.expo2010.cn). We take the amount of daily visitors as the observation of time series to study the visitor flow pattern of Expo 2010. During the exhibition period of 184 days, there are approximately 73,091,000 visitors in total and 397,234 visitors on average. Figure 1 below exhibits the general fluctuation pattern within which the horizontal axis represents the time (day) and the left and right vertical axes represent daily and accumulative visitor volume of Expo 2010 respectively.

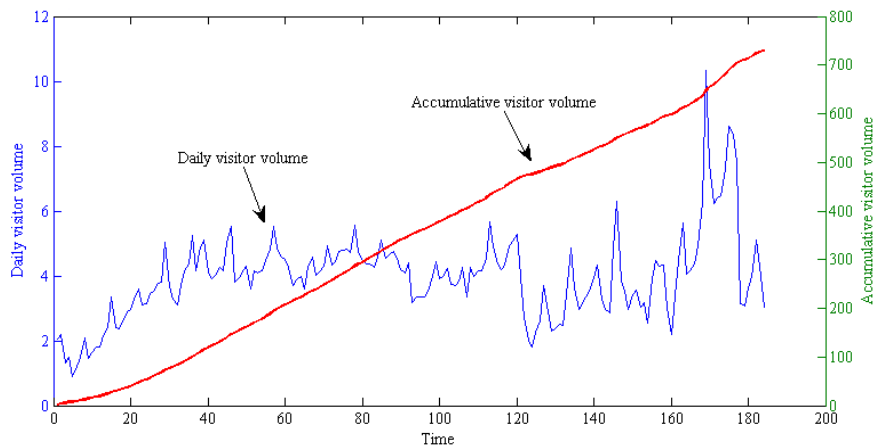

Fig.1: (Color on line) Daily and accumulative visitor volume ($\times 10^5$) of Expo 2010

## 3. The general feature of Expo visitor flow

The Hurst exponent, namely, long-range correlation exponent is used as a measure of the long term memory of time series, i.e. the autocorrelation of the time series, and it has been extensively used in many fields such as the stock market. The closer $H$ gets to 0.5, the more noise and fluctuation there will be in the time series. Meanwhile, when $H$ deviates more from 0.5 and tends towards 1, the time series will be more regular and persistent. Conversely, the time series are deemed to be anti-persistent when $H$ clines to 0.

In this section, we use the method of Rescaled Range Analysis [11, 12] to obtain the Hurst exponent of time series. The result is that the Hurst exponent of whole period daily visitor flow is 0.856 with $R^2 = 0.985$. It can be concluded that the visitor flow of Expo don't obey random walk but exhibit long-term stability and correlation due to that the Hurst exponent surpasses 0.5. The deviations of the visitor flow in the future tend to keep the same sign like the past.

Table 1 Mean value and standard deviation ($\times 10^3$) of visitor volume of Expo 2010

| Period | May. | Jun. | Jul. | Aug. | Sept. | Oct. | Total |
|---|---|---|---|---|---|---|---|
| Mean value | 259.2 | 436.9 | 444.8 | 401.9 | 333.6 | 506.3 | 397.2 |
| Standard deviation | 96.0 | 55.1 | 40.3 | 73.4 | 89.2 | 194.7 | 132.0 |

As shown in Table 1, we also calculate the mean value and standard deviation (with thousand visitors) of the monthly and whole visitor volume. From the results we can see the general fluctuation pattern of monthly visitor flow. Firstly, much fewer people come to Expo in May owing to that people think they have enough time to visit, thus there is no need to hurry. Besides, the operation of Expo is not in the best condition which makes people hesitate to visit. Secondly, the visitor flow in July is the most stationary and the second greatest which are the results of steady flow of tour groups and students who are in their summer vocation. Thirdly, during the six months of exhibition, October has the greatest visitor volume and standard deviation. The outbreak of visitor volume at the end of exhibition is not surprising. Meanwhile, the obvious fluctuation in visitor flow is the result of interweaving of small visitor volume on the 14 designated days and extremely great visitor volume on several standard days such as 1,032,700 on 16th, 859,900 on 22nd and 837,400 on 23rd.

In a word, the general law of Expo visitor flow exhibits dual features of both fluctuation and stability. The daily volumes are correlated from the long-term perspective, nevertheless show obvious fluctuation in short period.

**4. Statistical properties of visibility graph of Expo visitor volume**

According to the algorithm [6], a visibility graph is obtained from the mapping of a time series into a network according to the following visibility criterion: two arbitrary data $(t_a, y_a)$ and $(t_b, y_b)$ in the time series have visibility and consequently become two connected nodes in the associated graph, if any other data $(t_c, y_c)$ such that $t_a < t_c < t_b$ fulfills:

$$y_c < y_a + (y_b - y_a)\frac{t_c - t_a}{t_c - t_a} \qquad (4)$$

Figure 2 shows a typical example of this algorithm. In the upper panel, the data are displayed as vertical bars with heights indicating the values and the visibilities between data points are expressed as dashed lines. The converted network is shown in the lower panel, where the nodes correspond to series data in the same order and an edge connects two nodes if there is visibility between them. The visibility graphs inherit several properties of time series in its structure. More specifically, periodic series convert into regular graphs, random series do so into exponential random graphs, and fractal series convert into scale-free networks.

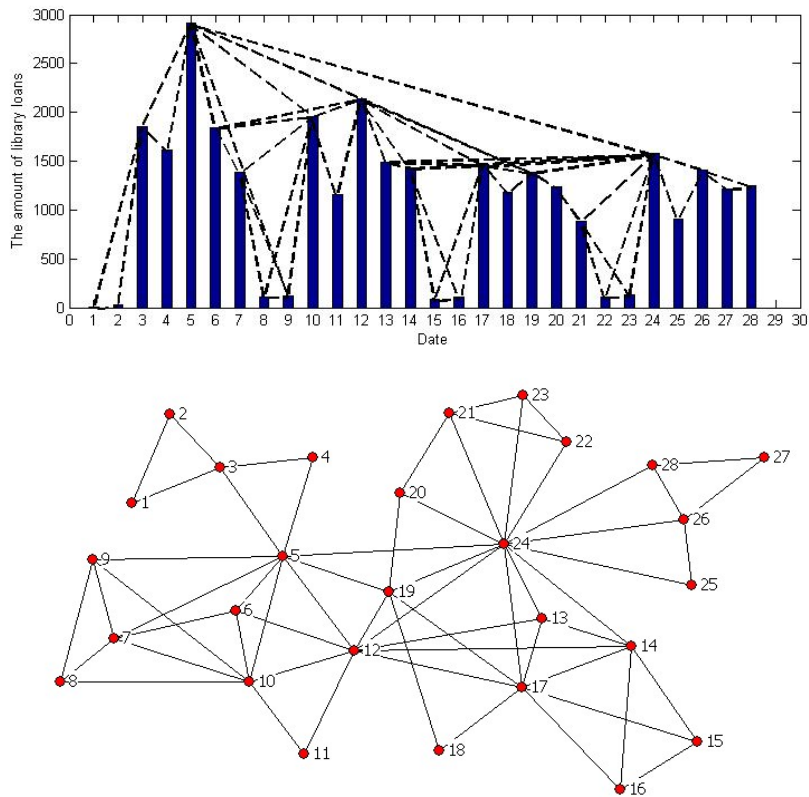

Fig.2: (Color on line) A typical example of visibility graph algorithm.

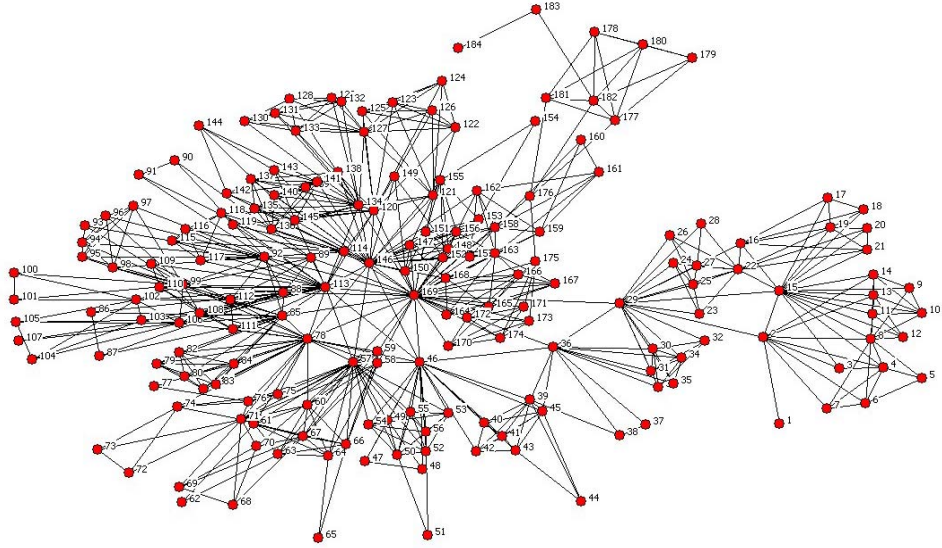

Fig.3: (Color on line) The network mapped from time series of daily visitor volume of Expo 2010.

As shown in Fig.3, the time series of daily visitor volume of Expo have been converted to visibility graphs using the algorithm introduced above. In order to investigate the feature of original time series of visitor flow, some characteristics of the visibility graph with 184 nodes and 676 edges were calculated as follows [2]:

1) Average degree of network $\langle k \rangle$: the mean value of degrees of all the nodes in network, $\langle k \rangle = \sum_{i=1}^{N} k_i / N = 7.348$.

2) Average clustering coefficient $\langle C \rangle$: the mean value of clustering coefficients of all the nodes in network, $\langle C \rangle = \sum_{i=1}^{N} C_i / N = 0.789$.

3) Diameter of network $D$: the maximal distance between any pair of nodes, $D = \max_{i,j} d_{ij} = 8$, where $d_{ij}$ refers to the number of edges on the shortest path connecting node $i$ and $j$.

4) Average path length $L$: the mean value of the distance of any pair of nodes, $L = \frac{2}{N(N+1)} \sum_{i \geq j} d_{ij} = 3.447$.

5) Degree distribution $P(k)$: the probability of a certain node to have degree $k$. It is

called scale-free network when degree distribution obeys a right-skewed power-law $P(k) \sim k^{-\gamma}$. For the sake of decreasing the noise, we study the accumulative degree distribution which behaves power-law as $P(k) = 44.25 k^{-2.37}$ with $R^2 = 0.991$. Therefore, the visibility graph is scale-free.

6) The small-world effect: calculating the average path length step by step while increasing the total number of nodes $N$ in network. If $L$ increases logarithmically along with $N$, namely, $L(N) \sim \ln N$, or more slower, while the network keeping a high clustering coefficient, then the network is regarded with the feature of small world. As shown in Fig.4, it can be observed that $L$ increases along with $N$ slower than logarithmical pattern. Consequently, the visibility graph is a small-world network.

7) Hierarchical structure: weighted average values of clustering coefficients of nodes with degree $k$ were calculated as $\overline{C}(k) = \langle C | k \rangle = \sum_{i=1}^{n} C_i / n$, where $n$ refers to the number of different clustering coefficients a node with degree $k$ has. The network is considered to be hierarchically constructed if $\overline{C}(k) \sim k^{-\alpha}$. In our case, $\overline{C}(k) = 5.597 k^{-0.97}$.

8) Pearson correlation coefficient $r$ [13]: there are many hub-nodes in scale-free network who own much greater degree than other nodes. Whether the interaction among the hubs of the network is attraction or repulsion can be determined from the correlation between the degrees of different nodes. The degree correlation can be quantified by Pearson correlation coefficient defined as:

$$r = \frac{N^{-1} \sum_i k_1 k_2 - \left[ N^{-1} \sum_i \frac{1}{2}(k_1 + k_2) \right]^2}{N^{-1} \sum_i \frac{1}{2}(k_1^2 + k_2^2) - \left[ N^{-1} \sum_i \frac{1}{2}(k_1 + k_2) \right]^2},$$

where $k_1$ and $k_2$ are the degrees of the nodes at the two ends of edge $i$. It can be deduced from the result $r = 0.115$ that the network is positively correlated with hub-nodes being attracted to each other.

9) Nearest neighbors average connectivity [14]: the relation of degree between one node and its nearest neighbors can be measured by the quantity $\langle K_{nn} \rangle = \sum_{k'} k' P(k'|k)$, where the conditional probability $P(k'|k)$ denotes that a node with degree $k$ is connected to a node with degree $k'$. Therefore, $\langle K_{nn} \rangle$ is used to investigate the relation between the degree of a certain node and the average degree of its nearest neighbors. From Fig.4 we can see clearly that $\langle K_{nn} \rangle$ and $k$ are positively correlated.

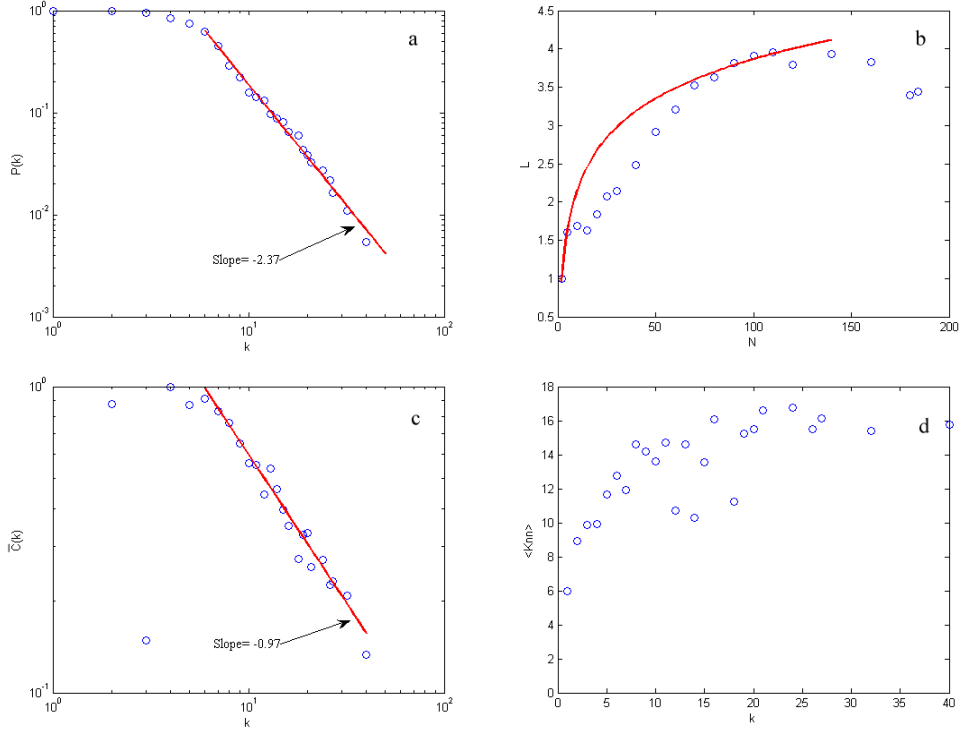

Fig.4 (Color online) The topological feature of the visitor flow network (a. The accumulative degree distribution behaves power-law with exponent -2.37. b. The average path length grows with the total number of nodes in network slower than logarithmical pattern. c. The weighted average value of clustering coefficients decreases with degree as power-law with exponent near to 1. d. The plot of $\langle K_{nn} \rangle \sim k$ shows the relation of degree between a node and its neighbors is positively correlated.)

From the topological parameters calculated above, some conclusions may be obtained as follows:

Firstly, the visibility graph being scale-free networks verifies that the time series of Expo visitor flow to be with fractal characteristics, enhancing the fact that power-law degree distributions are related to fractality, something highly discussed recently [6-10, 15-18]. More accurately, the total degree of the top 29 of 184 nodes with degree no less than 10 accounts for up to 36.1% of the sum degree of the whole network, and the average degree of the remaining 155 nodes is only 5.574.

Secondly, the facts that the network owns high clustering coefficient and low average path length which grow slowly with the total number of nodes verified the small-world phenomenon which means there are tight connections between nodes even they are located far away from each other for there are visibility lines between the corresponding data points in time series. Consequently, it can be deduced that some certain relations exist between the amounts of visitor flow in different time of whole exhibition. In another words, it is not random or unconnected between the past and the future in time series of human behaviours.

Thirdly, it can be inferred from the result $\overline{C}(k) = 5.597 k^{-0.97}$ that the visibility graph is hierarchically constructed which means if a node in network has greater degree, its neighbors are not tended to be connected with each other. The nodes with greater degrees are the ones own relatively much greater observations in time series than their directly connected and even unconnected neighbors. These extreme points correspond to the hub-nodes in scale-free networks. For example, node 15 has comparatively high degree $k = 18$ and low clustering coefficient $C = 0.261$, and there are 335,300 visitors on May 15th, much greater than its neighbors (There are 163,000, 180,400, 180,100, 215,500, 240,300 visitors on the 5 days before the 15th respectively and 241,500, 236,400, 261,900, 290,600, 296,400 visitors after the 15th. Obviously, it is much lower and more uniform before and after the 15th.). Correspondingly, node 15 has many neighbors in the network who are separated into two parts and connected with each other with small probability. Thus, it can be concluded that it is inevitable to show some extreme visitor volume in homogeneous flow on such pattern,

Finally, the result $r = 0.115$ means the network is assortative mixing, within which the nodes with high degrees tend to link to the nodes also have high degrees. Moreover, the fact $\langle K_{nn} \rangle$ and $r$ are positively correlated implies that the greater degree of a certain node, the greater average degree of its neighbors. To make the problem more clearly, we studied the relation between the node degree in network and the average visitor volume in time series. As shown in Fig.5, generally, the nodes with greater degree correspond to the data point with higher visitor volume. To sum up, the clustering phenomenon of hub-nodes in network means extreme points in time series appear in the form of group, in another word, large visitor flow always accompany with other large visitor flows. Consequently, there are some consecutive periods of extreme visitor volumes in the whole exhibition, for instance, from Oct.14th to Oct.24th with 726,255 visitors on average and from Jul.10th to Jul.28th with 466,942 daily visitors.

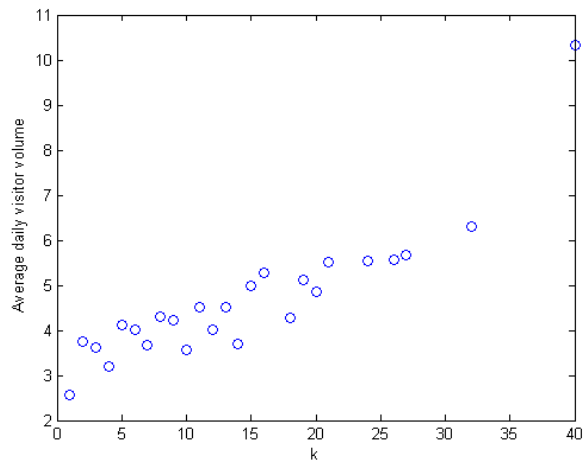

Fig.5 (Color online) The relation between node degree and visitor volume ($\times 10^5$)

In conclusion, the visibility graphs converted from time series are scale-free, small-world and hierarchically constructed. Thus the original time series are deemed to be with fractal feature and there are close relationships between the data points, especially those extreme points. Furthermore, the existence of extreme visitor volume is unavoidable and consecutive.

## 5. Summary and discussion

Expo 2010 Shanghai China is a wonderful event which has grand scale, long exhibition period and huge visitor amount, remaining us with the best memories and valuable experiences. In our research, the visitor flow is investigated from two different viewpoints:

On one hand, we discuss the general feature of the daily visitor volume from the perspective of statistics and time series analysis. The results of Hurst exponent, mean value and standard deviation verify that the Expo visitor flow shows mixing properties of stability in long-term and fluctuation in short-term period.

On the other hand, the complex network converted from time series with visibility algorithm exhibits scale-free property, small-world effect and hierarchical structure, which confirm that the original time series are fractal within which data points are intensively connected with each other. Furthermore, the relations between degree of one node and its neighbors, degree and clustering coefficient as well as degree and visitor volume prove that the existence of extreme visitor volume is inevitable and appear in group.

The method and conclusion of our work may be helpful to manage and forecast the visitor flow of such large-scale exhibitions and spots meeting or other human repetitious behaviours on collective level, as well as enrich the researches of the correlation between time series and complex networks.

**Acknowledgement**

This paper is supported by the National Natural Science Foundation of China (Grant No.70871082) and the Foundation of Shanghai Leading Academic Discipline Project (Grant No.S30504).